\documentclass{article}
\usepackage{amssymb,epsfig}
\title{Quark deconfinement in neutron star cores and the ground state of neutral matter}
\author{Chang-Qun Ma and Chun-Yuan Gao\footnote{Electronic address:
gaocy@pku.edu.cn}\\
School of Physics, Peking University, Beijing 100871, China}
\begin{document}
\maketitle
\begin{abstract}
Whether or not deconfined quark phase exists in neutron star cores
and represents the ground state of neutral matter at moderate
densities are open questions. We use two realistic effective quark
models, the three-flavor Nambu-Jona-Lasinio model and the modified
quark-meson coupling model, to describe the neutron star matter.
After constructing possible hybrid equations of state (EOSes) with
unpaired or color superconducting quark phase, we systematically
discuss the observational constraints of neutron stars on the EOSes.
It is found that the neutron star with pure quark matter core is
unstable and the hadronic phase with hyperons is denied, while
hybrid EOSes with two-flavor color superconducting phase or unpaired
quark matter phase are both allowed by the tight and most reliable
constraints from two stars Ter 5 I and EXO 0748-676. And the hybrid
EOS with unpaired quark matter phase is allowed even compared with
the tightest constraint from the most massive pulsar star PSR
J0751+1807. Therefore, we conclude that the ground state of neutral
matter at moderate densities is in deconfined quark phase likely.

\bigskip
\noindent PACS number(s): 12.38.Mh, 12.39.-x, 26.60.+c
\end{abstract}
\section{INTRODUCTION}\label{intro}
Neutron stars are some of the densest objects in the universe and
the density in the inner core of a neutron star could be as large as
several times nuclear saturation density ($\cong0.17$ fm$^{-3}$)
\cite{l}. The core of neutron star is so dense that phase transition
from confined hadronic phase to deconfined quark phase may exist.
The possible emergence of deconfinement phase in neutron star cores
has aroused great interests since it may have a distinct effect on
the neutron star structure\cite{dec_in_star}. Up to now, however, no
conclusive observational or experimental evidence suggests that the
quark matter core conjecture is true and this still remains an open
question. In our recent work where the hadronic phase containing
octet baryons was considered, it was found that the equation of
state (EOS) with deconfinement phase would be ruled out by the
observational mass limit of 1.68M$_{\odot}$ of star Ter 5
I\cite{mg}. Similar result was also educed by \"Ozel who reported
that EOSes with exotic phases would be ruled out by the inferred
mass and radius for star EXO 0748-676 and she concluded that the
ground state of matter was hadrons and not deconfined
quarks\cite{o}. While Alford et al. compared \"Ozel's observational
limits with predictions based on a more comprehensive set of
proposed equations of state from the literature, and concluded that
the presence of quark matter in EXO 0748-676 was not ruled
out\cite{AB}. Therefore, the existence of deconfined quark phase in
neutron star cores and the ground state of neutral matter at
moderate densities are controversial. Our goal of the present paper
is to systematically investigate the observational constraints on
the deconfinement in neutron star cores and check whether deconfined
quarks are possible to exist in the ground state of neutral matter
at moderate densities.

It is expected that at extreme conditions chiral symmetry can be
restored and quarks and gluons become deconfined\cite{qgp}.
Consequently, quark matter was frequently dealt with as a
non-interacting quark gas, i.e., unpaired quark matter (UQM), and
usually described by a phenomenological bag model\cite{arkcv}.
According to the BCS theory\cite{bcs}, any attractive interaction in
a cold fermi sea will cause Cooper instability in the vicinity of
fermi surface in the momentum space and superconductor will be
formed. Because of the attractive quark-quark interaction in the
color antitriple channel\cite{attractive} it is expected that color
superconducting state, with a spontaneous breakdown of the
non-Abelian SU(3) color gauge group, would be the ground state of
quark matter. The color superconducting state has attracted great
interests since it was found that superconducting gap could be
$\sim$ 100MeV due to the nonperturbative features\cite{gap100}.
Depending on quarks participating in a diquark condensation, one can
distinguish several color superconducting phases. The prominent two
are the two-flavor color superconducting (2SC) phase containing $ud$
pairs together with unpaired strange quarks and the color-flavor
locked (CFL) phase containing $ud$, $ds$ and $us$ pairs(for recent
reviews see Ref. \cite{csreview}).

We adopt two realistic effective quark models for quark and hadronic
phases respectively to describe the neutron star matter. The
extended three-flavor Nambu-Jona-Lasinio (NJL) model with
determinant interaction\cite{njlreview}, which shares many
symmetries with QCD, is used to calculate the properties of quark
matter in UQM, 2SC and CFL phases. For hadronic phase, we adopt the
EOSes of neutron-proton matter and hypernuclear matter with octet
baryons by the improved modified quark-meson coupling (MQMC)
model\cite{mg,ph}. The MQMC model gives satisfactory description for
saturation properties of nuclear matter\cite{st} and could reproduce
the bulk properties of finite nuclei well\cite{gsr}. We then assume
a sharp (first order) transition from pure hadronic to pure quark
phase and consider only the homogeneous quark phase and not mixed
phase between different quark phases.

The outline of this paper is as follows: In Sec. \ref{sec2}, we
briefly introduce the three-flavor NJL model for dense quark matter
and the MQMC model for hadronic phase. In Sec. \ref{sec3}, we
construct the possible equations of state of neutron star matter and
then discuss the observational constraints on the EOSes. Sec.
\ref{sec4} is devoted to summaries and conclusions.

\section{THE MODEL}\label{sec2}
\subsection{QUARK PHASE}
Strange quark matter at moderate densities can be effectively
described by the three-flavor NJL model\cite{fmm,bo}. The Lagrange
density of extended three-flavor NJL model with six-fermion
determinant interaction (t'Hooft term) is given by
\begin{equation}
{\cal L}_{\rm{NJL}}=\bar{q}\left({\rm
i}\gamma^\mu\partial_\mu-\hat{m}_0\right)q+{\cal L}_{\bar{q}q}+{\cal
L}_{qq},
\end{equation}
where
\begin{eqnarray}
{\cal L}_{\bar{q}q}=\!\!\!\!&&
G\sum\limits_{a=0}^{8}\left[\left(\bar{q}\tau_{a}q\right)^2+\left(\bar{q}{\rm
i}\gamma_5\tau_aq\right)^2\right]
\nonumber\\
&&-K\left[{\rm{det}}_{f}\left(\bar{q}(1+\gamma_5)q\right)+{\rm{det}}_f\left(\bar{q}(1-\gamma_5)q\right)\right]
\end{eqnarray}
and
\begin{equation}{\cal
L}_{qq}=H\sum\limits_{A=2,5,7}\sum\limits_{A'=2,5,7}\left(\bar{q}{\rm
i}\gamma_5\tau_A
\lambda_{A'}q^c\right)\left(\bar{q}^ci\gamma_5\tau_A\lambda_{A'}q\right).
\end{equation}
Here, $q=(u,d,s)^T$ denotes the quark fields with three colors. And
the current quark mass matrix has the form
$\hat{m}_0={\rm{diag}}(m_{0u},m_{0d},m_{0s})$ in the flavor space,
where $m_{0u}=m_{0d}=m_{0q}$ is assumed throughout this paper.
$\tau_0$=$\sqrt{\frac{2}{3}}\mathbb{1}$ is proportional to the unit
matrix in the flavor space. $\tau_A$ and $\lambda_A$ (A=1,\ldots,8)
are Gell-Mann matrixes in flavor and color spaces respectively.
$q^c=C\bar{q}^T$ is the charge-conjugate spinor.

In the present work, we restrict ourselves to bulk quark matter in
mean-field approximation and focus on the chiral condensates defined
as
\begin{equation}
\phi_f=\left<\bar{q}_fq_f\right>,\hskip5mm f=u,d,s,
\end{equation}
and the three-flavor diquark condensates being
\begin{equation}
\Delta_A=-2H\left<\bar{q}^c\gamma_5\tau_A\lambda_Aq\right>,
\hskip5mmA=2,5,7.
\end{equation}
After bosonization, one obtains the linearized version of the model
in the mean-field approximation,
\begin{eqnarray}
{\cal
L}=\!\!\!\!&&\bar{q}\left({\rm i}\gamma^{\mu}\partial_{\mu}-\hat{m}\right)q\nonumber\\
&&+\frac{1}{2}\sum\limits_A\left[\bar{q}\left(\Delta_A\gamma_5\tau_A\lambda_A\right)q^c
+\bar{q}^c\left(-\Delta_A^{\ast}\gamma_5\tau_A\lambda_A\right)q\right]\nonumber\\
&&-\frac{1}{4H}\sum\limits_A\left|\Delta_A\right|^2-2G\sum\limits_f\phi_f^2+4K\phi_u\phi_d\phi_s,
\end{eqnarray}
where we have introduced the constituent quark mass
\begin{eqnarray}
\hat{m}=\left(
                    \begin{array}{ccc}
                      \displaystyle m_{0u}-4G\phi_u+2K\phi_d\phi_s&&\\
                       &\displaystyle m_{0d}-4G\phi_d-2K\phi_s\phi_u&\\
                       &&\displaystyle m_{0s}-4G\phi_s-2K\phi_u\phi_d
                    \end{array}
                  \right).
\end{eqnarray}
Employing Nambu-Gorkov formalism, then the thermodynamic potential
per unit volume at temperature $T$ is obtained via the finite
temperature field theory\cite{field} and takes the form,
\begin{equation}\label{potential}
\Omega=-\frac{T}{2V}\sum\limits_{\vec{P}}\sum\limits_{i=1}^{72}
\left[\frac{\left|\omega_i\right|}{2T}
+{\rm{ln}}\left(1+e^{-\left|\omega_i\right|/T}\right)\right]+\Omega_{\rm{const}}+\Omega_e,
\end{equation}
where
\begin{equation}
\Omega_{\rm{const}}=2G\sum\limits_{f=u,d,s}{\phi_f}^2
-4K{\phi_u}{\phi_d}{\phi_s}+\frac{1}{4H}\sum\limits_A\left|\Delta_A\right|^2,
\end{equation}
and
\begin{equation}
\Omega_e=-\frac{1}{12\pi^2}\left(\mu_e^4+2\pi^2T^2\mu_e^2+\frac{7\pi^4}{15}T^4\right)
\end{equation}
is the contribution from the electron gas with chemical potential
$\mu_e$.

In Eq.(\ref{potential}), $\omega_i$ is the energy of a quasiparticle
and can be obtained by diagonalizing the inverse propagator in
Nambu-Gorkov basis. Or, equivalently, $\omega_i$ can be obtained by
calculating the eigenvalues of the 72$\times$72 matrix
\begin{equation}
{\cal M}=\left[
        \begin{array}{cc}
          \displaystyle\vec{P}\cdot\vec{\alpha}+\hat{m}\gamma_0-\hat{\mu}&
           \displaystyle-\sum\nolimits_A\Delta_A\gamma_0\gamma_5\tau_A\lambda_A\\
          \displaystyle\sum\nolimits_A\Delta_A^\ast\gamma_0\gamma_5\tau_A\lambda_A
            &\displaystyle\vec{P}\cdot\vec{\alpha}+\hat{m}\gamma_0+\hat{\mu}
        \end{array}
      \right].
\end{equation}
And the chemical potential operator $\hat{\mu}$ is a diagonal
9$\times$9 matrix in flavor and color space. By introducing the
quark number chemical potential $\mu$, electrochemical potential
$\mu_Q$ and two additional chemical potentials $\mu_3$ and $\mu_8$
coupled to the color charges $\lambda_3$ and $\lambda_8$
respectively, $\hat{\mu}$ can be expressed as(see Ref. \cite{fmm}
for details)
\begin{equation}
\hat{\mu}=\mu+\mu_Q\left(\frac{1}{2}\tau_3+\frac{1}{2\sqrt{3}}\tau_8\right)
+\mu_3\lambda_3+\mu_8\lambda_8.
\end{equation}
In beta equilibrium, we have
\begin{equation}
\mu_e=-\mu_Q.
\end{equation}

The order parameters, $\phi_f$ and $\Delta_A$, can be obtained by
minimizing the thermodynamic potential, and are equivalently given
by the gap equations,
\begin{eqnarray}
&&\frac{\partial\Omega}{\partial\phi_f}=0, \hskip6mmf=u,d,s,\\
&&\frac{\partial\Omega}{\partial\Delta_A}=0, \hskip5mmA=2,5,7.
\end{eqnarray}
With the diquark condensates, we can explicitly distinguish
different quark phases as:
\begin{eqnarray}&&\Delta_2=\Delta_5=\Delta_7=0:\hskip5mm {\rm UQM};\nonumber\\
&&\Delta_5=\Delta_7=0, \Delta_2\neq0:\hskip4mm {\rm 2SC};\nonumber\\
&&\Delta_2, \Delta_5, \Delta_7\neq0:\hskip5mm\hskip5mm \hskip3mm{\rm
CFL}. \nonumber\end{eqnarray}

Dense quark matter in neutron star is electrical and color neutral,
then the following relations should be maintained:
\begin{eqnarray}
&&n_Q=-\frac{\partial\Omega}{\partial\mu_Q}=0,\\
&&n_3=-\frac{\partial\Omega}{\partial\mu_3}=0,\\
&&n_8=-\frac{\partial\Omega}{\partial\mu_8}=0.
\end{eqnarray}

Finally, for a given quark number density $$
n=-\frac{\partial\Omega}{\partial\mu},$$the energy density ${\cal
E}$ and pressure ${\cal P}$ at zero temperature are given by:
\begin{eqnarray}
&&{\cal E}=\Omega+\Omega_{\rm{vac}}+{\mu}n,\\
&&{\cal P}=-\Omega-\Omega_{\rm{vac}}.
\end{eqnarray}
$\Omega_{\rm{vac}}$ is chosen so that ${\cal P}$ and ${\cal E}$
vanish in vacuum, which is the only way to uniquely determine the
EOS within the NJL model without any further assumption.

For the model parameters, we take the values as follows. To
regularize the divergent integrals we need a sharp cutoff $\Lambda$
in 3-momentum space since the NJL model is nonrenormalizable. Thus
we have a total of 6 parameters, namely, the current masses $m_{0s}$
and $m_{0q}$ for strange and nonstrange quarks, the three couplings
$G$, $K$ and $H$, and the cutoff $\Lambda$. Following the method
adopted in Ref.\cite{parameters}, we get $\Lambda$=602.8MeV,
$G$$\Lambda^{2}$=1.803, $K$$\Lambda^{5}$=12.93 and $m_{0s}$=140.9MeV
by fitting the meson masses\cite{mass} $m_\pi=134.98$MeV,
$m_K=497.65$MeV and $m_{\eta'}=957.78$MeV and the $\pi$ decay
constant $f_{\pi}=92.2$MeV\cite{pidecay} while $m_{0q}$ is fixed at
5.5MeV. $H=G$ is set as has been used in Ref.\cite{fmm}.

\subsection{HADRONIC PHASE}
For the hadronic phases, we focus on the neutron-proton matter and
the hypernuclear matter consisting of the baryon octet, i.e., p, n,
$\Lambda$, $\Sigma^+$, $\Sigma^0$, $\Sigma^-$, $\Xi^0$ and $\Xi^-$.
The hadronic phases are described with the improved MQMC model which
has been discussed in detail in Ref.\cite{mg}. In order to make the
calculation for hadronic phases compatible with that for quark
phases, we should take the same current quark masses in both models.
Therefore the current quark masses take $m_{0u}=m_{0d}=5.5$MeV and
$m_{0s}$=140.9MeV, which are different from those used in
Ref.\cite{mg}, and the other inputs, such as the mass spectrum and
the nucleon's radius, are unchanged. As a consequence, the bag
constant in vacuum is changed to be $B_0^{1/4}$=187.7716MeV. The
obtained zero-point motion parameters and bag-radii for baryons and
coupling constants are all changed, their new values are listed in
TABLE \ref{tab1}. and TABLE \ref{tab2}, respectively.
\begin{table}
\caption{The new zero-point motion parameters and radii for the
modified quark-meson coupling model.}\label{tab1}
\begin{center}\begin{tabular}{c|ccc}\hline\hline
&M(MeV)&Z&\hskip1.5mmR(fm)\\
\hline
 N & 939.0  &  2.0403 & \hskip1.5mm0.6000 \\
 $\Lambda$ & 1115.7  & 1.8099  &\hskip1.5mm 0.6459 \\
 $\Sigma^{+}$ & 1189.4  & 1.6318  & \hskip1.5mm0.6722 \\
 $\Sigma^{0}$ & 1192.6  & 1.6236 & \hskip1.5mm0.6733 \\
 $\Sigma^{-}$ & 1197.4  & 1.6114 & \hskip1.5mm0.6749 \\
 $\Xi^{0}$ &  1314.8 & 1.4743  & \hskip1.5mm0.6922 \\
 $\Xi^{-}$ & 1321.3  &  1.4567 & \hskip1.5mm0.6942 \\
\hline \hline
\end{tabular}
\end{center}\end{table}
\begin{table}\caption{New independent coupling constants,
which have been defined in Ref.\cite{mg}.}\label{tab2}
\begin{center}\begin{tabular}{c|c|c|c} \hline\hline
\hskip2mm$g_{\sigma}^{u,d}$\hskip2mm&\hskip2mm $g_{\omega}^{u,d}$ \hskip2mm &
\hskip2mm $g_{\rho}^{u,d}$\hskip2mm & \hskip2mm$g_{\sigma}^{{\rm bag},N}$\hskip2mm \\
\hline
\hskip2mm0.9685\hskip2mm&\hskip2mm2.7071\hskip2mm&\hskip2mm7.9288\hskip2mm&
\hskip2mm6.8732\hskip2mm\\
\hline \hline
\end{tabular}\end{center}\end{table}
\section{RESULTS AND DISCUSSIONS}\label{sec3}
\subsection{EQUATION OF STATE}\label{eos}
\begin{figure} \centering
{\epsfig{file=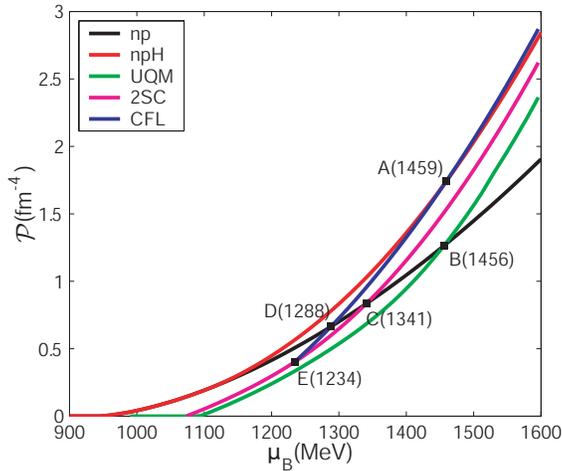,width=7.4cm}} \caption{Pressure of
hadronic or quark phase as a function of baryon chemical potential.
np (black) denotes neutron-proton matter and npH (red) denotes the
hypernuclear matter consisting of octet baryons. UQM (green), 2SC
(magenta) and CFL (blue) denote the three kinds of homogeneous quark
phases. Squares ordered by A, B, C and D mark the hadron-quark phase
transitions and E is the phase transition between 2SC and CFL
phases. Values in parentheses are the critical chemical
potentials.}\label{fig1}
\end{figure}

First order phase transition takes place at critical point where
baryon chemical potentials and pressures for the two phases are
equal. The pressures of hadronic and quark phases are shown in
Figure \ref{fig1}
as a function of baryon chemical potential and the position of
hadron-quark phase transition can be easily read off as the point
where the lines ${\cal P}(\mu_{B})$ cross. From the figure, we can
find four possible hadron-quark phase transitions, i.e., A, B, C and
D. EOS of the hypernuclear matter(npH) has no cross with those of
quark phases 2SC or UQM, and has only one cross at A with that of
CFL. Whereas the neutron-proton matter (np) could change into
deconfined quark phases in UQM, 2SC or CFL state at B, C and D.
Therefore, we could construct totally four kinds of hybrid EOSes
between hadronic phase and quark deconfinement phase.

\begin{figure}
\centering{\epsfig{file=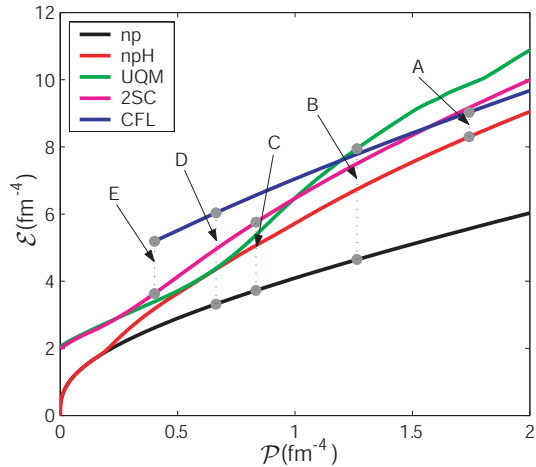,width=7cm}} \caption{Energy vs.
pressure for equations of state. The solid lines represent the EOSes
of pure phases. And the thin dotted lines indicate the first order
phase transitions. A, B, C, D and E are the same as those defined in
Figure \ref{fig1}.}\label{fig2}
\end{figure}

EOSes are plotted in Figure \ref{fig2} and possible phase transitions
are also shown by dotted lines marked by A, B, C, D and E. The
hybrid EOS of npH+CFL can be constructed by combining the EOS of
pure hypernuclear matter npH and the EOS of pure quark matter in CFL
state for low and high pressures respectively, with a hadron-quark
phase transition at A. The other EOSes of np+UQM, np+2SC and np+CFL
could be similarly constructed. There are some notable features for
the hybrid EOSes. First, the hadron-quark phase transitions are
first order, which is an inevitable result because we have chosen a
sharp interface as claimed in Sec. \ref{intro}. Second, the energy
gaps of phase transitions are rather large, which is found to have
profound effect on the star structure and is going to be discussed.
Among the hybrid EOSes, the stiffest one is of np+UQM, and the next
are of np+2SC, np+CFL and npH+CFL in turn.

\subsection{MAXIMUM MASS}
\begin{figure}
\centering{\epsfig{file=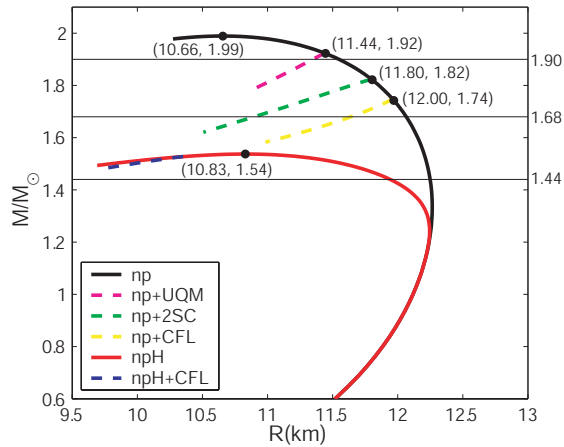,width=7.4cm}} \caption{Neutron star
mass-radius relations for pure or hybrid EOSes. Solid lines indicate
the pure hadronic star and dashed lines are for hybrid stars. The
dots denote the maximum masses and their coordinates are given in
the parentheses. Three horizontal lines represent the observational
constraints from stars PSR 1913$+$16
(1.44M$_{\odot}$\cite{psr1913}), Terzan 5 I
(1.68M$_{\odot}$\cite{ter}) and PSR J0751+1807
(1.90M$_{\odot}$\cite{psrj051}) respectively.}\label{fig3}
\end{figure}

By solving the Tolman-Oppenheimer-Volkoff equations \cite{tov},
mass-radius relations are obtained for different equations of state,
which are given in Figure \ref{fig3}. The softest EOS of npH+CFL
predicts a maximum mass of 1.54 M$_{\odot}$. Compared with the best
measured pulsar mass 1.44M$_\odot$ in the binary pulsar PSR
1913$+$16\cite{psr1913}, which had been taken as the lower limit of
neutron star's maximum mass for many years, all the EOSes here are
allowed. However, very recent measurements strongly indicate that
there are more massive neutron stars. The typical neutron star is
reported by Ransom et al. who inferred that at least one of the
stars in Terzan 5, the Ter 5 I, is more massive than 1.48, 1.68, or
1.74 M$_{\odot}$ at 99\%, 95\%, and 90\% confidence
levels\cite{ter}. Therefore, as indicated in Figure
\ref{fig3},
imposing the tighter observational constraint of 1.68M$_{\odot}$ at
95\% confidence level, pure hadronic EOS with hyperons and the
hybrid EOS of npH+CFL are firmly ruled out, while the hybrid EOSes
of np+UQM, np+2SC and np+CFL can be compatible with this constraint.
So far, the most massive pulsar star reported is PSR J0751+1807,
which has a inferred mass of 2.1$\pm$0.2M$_{\odot}$ covering 68\%
confidence uncertainties\cite{psrj051}. Therefore, PSR J0751+1807
gives the tightest mass constraint of 1.9M$_{\odot}$ at about 68\%
confidence level. If this tightest mass limit is confirmed, all the
hybrid EOSes are ruled out but that of np+UQM. And, of course, the
EOS of pure neutron-proton matter is allowed because it is stiffer
than the one of np+UQM. However, the measurement uncertainty of PSR
J0751+1807 is not yet small enough to draw any firm conclusion and
the tight constraint coming from Ter 5 I is more reliable.

The hybrid stars with quark phases are indicated in Figure
\ref{fig3} with dashed lines, which reveal that all the stars with
quark phase cores are unstable, i.e., those with quark cores will
collapse. The direct reason is that hybrid EOS has a rather large
discontinuity of energy caused by the first order phase transition
as claimed in Sec. \ref{eos}. Our result confirms those obtained in
earlier works, where the hadronic phase has been described by
different models, e.g., the relativistic mean field model or the
microscopic many-body theory\cite{sbb,mgpsd}. Nevertheless Buballa
et al. reported recently that by adopting another parameter set
which was determined by a different method from here, 2SC phase was
possible to exist in the star core within a very tiny
window\cite{bn}. Baldo et al. suggested that the instability may be
linked to the lack of confinement in the current NJL
model\cite{mgpsd}. Moreover, parameters obtained by fitting the
vacuum properties might not be suitable at high densities. Therefore
if the existence of pure quark cores in neutron stars was confirmed
by future observations, the NJL model currently used should be
modified.
\subsection{GRAVITATIONAL REDSHIFT}
\begin{figure}
\centering{\epsfig{file=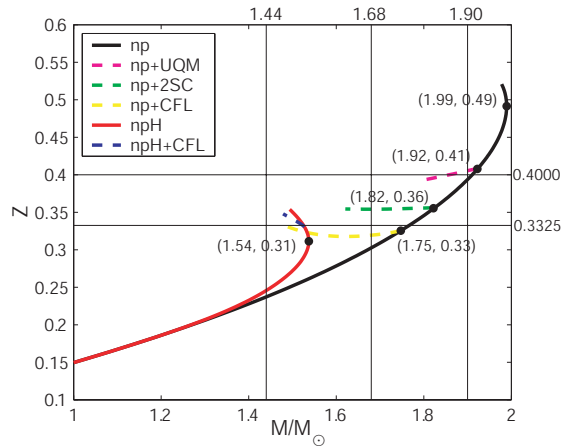,width=7.4cm}}
\caption{Gravitational redshifts vs. masses for different EOSes. The
lower horizontal line is the observational gravitational redshift of
EOX0748-676\cite{redshift,redshifterror} and the upper one is that
of 4U1700+24\cite{higherredshift}. The dots denote the maximum
masses and their coordinates are given in the parentheses.
Constraints of observational masses (the same as in Figure
\ref{fig3}) are also plotted with three vertical lines.}\label{fig4}
\end{figure}

In principle, the range of gravitational redshift predicted by an
EOS should cover all the ever detected redshifts. Cottam et al. have
reported a redshift $z$ of 0.35 inferred by identifying three sets
of transitions in the spectra of x-ray binary EXO
0748-676\cite{redshift}. The result has been confirmed by Chang et
al.\cite{redshiftcon}. And it has been shown that the total error in
this redshift is no more than 5\%\cite{redshifterror}. Therefore,
allowing for the error bar, it imposes a lower limit of about 0.3325
to the maximum redshift. In Figure \ref{fig4}, gravitational redshifts
vs. masses of different EOSes are shown. EOSes with hyperons fail to
construct stable neutron star satisfying the redshift of EXO
0748-676, so they are ruled out. Without hyperons, hybrid EOS of
np+CFL is not allowed likely, while those stiffer, namely, of
np+2SC, np+UQM and np, are permitted.

Larger observational redshift gives more stringent constraint.
Tiengo, et al., inferred that a redshift $z$=0.4 can be obtained by
explaining the emission lines from 4U1700+24 with the Ne IX
triplet\cite{higherredshift}. Then the hybrid EOS of np+2SC would be
ruled out by this constraint, and only the one of np+UQM is
marginally permitted. However, due to the lack of identifications of
other spectral features at $z$=0.4, it seems that the interpretation
of $z$=0.012 for 4U1700+24 is more reliable\cite{higherredshift}.
Therefore redshift of EXO 0748-676 is the most reliable constraint
till now.

Combining the constraints of observational masses and redshifts of
neutron stars, it can be summarized that EOSes of np+2SC, np+UQM and
np are very likely to be favored. And if either the mass limit of
1.9M$_{\odot}$ of the most massive pulsar star PSR J0751+1807 or the
very tentative interpretation of redshift of 4U1700+24 equal to 0.4
is confirmed, EOSes with CFL or 2SC quark phase are both denied,
then the allowed hybrid EOS is only of np+UQM. It should be noted
that rotation effect is not considered here because influences of
rotation effect on static properties are negligible for the main
observations\cite{rotation} we currently discussed.
\section{SUMMARIES AND CONCLUSIONS}\label{sec4}
We have used two realistic effective quark models, i.e., the
three-flavor NJL model and the MQMC model, to describe the neutron
star matter. For hadronic phase, EOSes with and without hyperons
(npH and np) are both considered. And we have discussed the quark
matter phase in normal and color superconducting states, namely,
UQM, 2SC and CFL. Then four possible hybrid EOSes between hadronic
phase and quark deconfinement phase are constructed. We find that
EOSes with hyperons should be ruled out by the observational
constraints from the mass of star Ter 5 I and/or the redshift of
binary star EXO 0748-676. Moreover, the hybrid EOS of np+CFL is also
ruled out by the observational redshift of EXO 0748-676. As a
consequence, hybrid EOSes of np+2SC and np+UQM as well as pure
hadronic EOS of np are most likely to be favored. Tightest but less
reliable constraint can be inferred by the mass of the most massive
pulsar star PSR J0751+1807, and if it is confirmed the permitted
hybrid EOS is of np+UQM only.

Therefore, we conclude that observational constraints of neutron
star could not rule out all the possible EOSes with quark phase
though the neutron stars with pure quark cores are found to be
unstable in our calculation. Both normal unpaired quark state and
two-flavor color superconducting state are likely permitted, and
future observations are needed to determine which of them (or none)
is the right ground state at this density. So the ground state of
neutral matter at moderate densities could be in deconfined quark
phase.

\section*{ACKNOWLEDGMENTS}
Financial support by the National Natural Science Foundation of
China under grants 10305001, 10475002 \& 10435080 is gratefully
acknowledged.


\begin{thebibliography}{20}
\bibitem{l}J. M. Lattimer and M. Prakash, Science, {\bf304}, 536 (2004).
\bibitem{dec_in_star}M. Alford and S. Reddy, Phys. Rev. D, {\bf67}, 074024
(2003).
\bibitem{mg}C. Q. Ma and C. Y. Gao, nucl-th/0612107v2 (2007).
\bibitem{o}F. $\ddot{\rm O}$zel, Nature, {\bf441}, 1115 (2006).
\bibitem{AB} M. Alford, D. Blaschke, A. Drago,
T. Kl\"ahn, G. Pagliara and J. Schaffner-Bielich, Nature, {\bf445},
E7 (2007).
\bibitem{qgp}J. C. Collins and M. J. Perry, Phys. Rev. Lett., {\bf34}, 1353
(1975); E. V. Shuryak, Phys. Lett. B, {\bf78}, 150 (1978).
\bibitem{arkcv}A. Chodos, R. L. Jaffe, K. Johnson, C. B. Thorn and V. F.
Weisskopf, Phys. Rev. D, {\bf9}, 3471 (1974); E. Farhi and R. L.
Jaffe, Phys. Rev. D, {\bf30}, 2379 (1984).
\bibitem{bcs}J. Bardeen, L. N. Cooper and J. R. Schrieffer, Phys. Rev., {\bf106},
162 (1957); Phys. Rev., {\bf108}, 1175 (1957).
\bibitem{attractive}B. Barrois, Nucl. Phys. B, {\bf129}, 390 (1977); D. Bailin and
A. Love, Phys. Rep., {\bf107}, 325 (1984).
\bibitem{gap100}R. Rapp, T. Sch\"afer, E. Shuryak and M. Velkovsky, Phys. Rev.
Lett., {\bf81}, 53 (1998); M. Alford, K. Rajagopal and F. Wilczek,
Phys. Lett. B, {\bf422}, 247 (1998).
\bibitem{csreview}D. H. Rischke, Prog. Part. Nucl. Phys., {\bf52}, 197 (2004);
\bibitem{njlreview}Y. Nambu and G. Jona-Lasinio, Phys. Rev., {\bf122}, 345 (1961);
{\bf124}, 246 (1961) and for reviews of the model see S. P.
Klevansky, Rev. Mod. Phys., {\bf64}, 649 (1992); T. Hatsuda and T.
Kunihiro, Phys. Rep., {\bf247}, 221 (1994); J. Bijnens, Phys. Rep.,
{\bf265}, 370 (1996).
\bibitem{ph}S. Pal, M. Hanauske, I. Zakout, H. St$\ddot{\rm o}$cker and
W. Greiner, Phys. Rev. C, {\bf60}, 015802 (1999).
\bibitem{st}H. M$\ddot{\rm u}$ller, B. K. Jennings, Nucl. Phys. A, {\bf626}, 966 (1997);
 J. C. Caillon and J. Labarsouque, Phys. Lett. B, {\bf425}, 13 (1998).
\bibitem{gsr}H. M$\ddot{\rm u}$ller, Phys. Rev. C, {\bf57}, 1974 (1998).
\bibitem{fmm}F. Neumann, M. Buballa and M. Oertel, Nucl. Phys. A, {\bf714}, 481 (2003).
\bibitem{bo}M. Buballa and M. Oertel, Phys. Lett. B, {\bf457}, 261
(1999); Nucl. Phys. A, {\bf703}, 770 (2002); A. W. Steiner, S. Reddy
and M. Prakash, Phys. Rev. D, {\bf66}, 094007 (2002); M. Buballa,
Phys. Rep., {\bf407}, 205 (2005) and references therein.
\bibitem{field}J. I. Kapusta, {\sl Finite-Temperature Field Theory} (Cambridge University
Press, Cambridge, England, 1989).
\bibitem{parameters}P. Rehberg, S. P. Klevansky and J. H\"ufner, Phys. Rev. C, {\bf53}, 410 (1996).
\bibitem{mass}W. M. Yao, et al.,  J. Phys. G: Nucl. Part. Phys., {\bf33}, 1
(2006).
\bibitem{pidecay}S. Descotes-Genon and B. Moussallam, Eur. Phys. J. C, {\bf42}, 403 (2005).
\bibitem{tov}R. C. Tolman, Phys. Rev., {\bf55}, 364 (1939);
J. R. Oppenheimer, G. Volkoff, Phys. Rev., {\bf55}, 374 (1939).
\bibitem{psr1913}J. M. Weisberg and J. H. Taylor, {\sl Radio Pulsars} ed M. Bailes,
D. J. Nice and S. Thorsett (San Francisco: Astronomical Society of
the Pacific) pp 93¨C8 (2003); S. E. Thorsett and D. Chakrabarty,
Astrophys. J., {\bf512}, 288 (1999).
\bibitem{ter}S. M. Ransom, J. W. T. Hessels and I. H. Stairs, et al., Science, {\bf307}, 892 (2005).
\bibitem{psrj051}D. J. Nice, E. M. Splaver and I. H. Stairs, et al., Astrophys. J., {\bf634}, 1242 (2005).
\bibitem{sbb}K. Schertler, S. Leupold and J. Schaffner-Bielich, Phys. Rev. C, {\bf60}, 025801
(1999); M. Baldo, M. Buballa and G. F. Burgio, et al., Phys. Lett.
B, {\bf562}, 153 (2003);
\bibitem{mgpsd}M. Baldo, G. F. Burgio, P. Castorina, S. Plumari and D. Zappal$\grave{\rm a}$,
Phys. Rev. C, {\bf75}, 035804 (2007).
\bibitem{bn}M. Buballa, F. Neumann, M. Oertel and I. Shovkovy, Phys. Lett. B, {\bf595}, 36 (2004).
\bibitem{redshift}J. Cottam, F. Paerels and M. Mendez, Nature, {\bf420}, 51 (2002).
\bibitem{redshiftcon}P. Chang, S. Morsink, L. Bildsten and I. Wasserman, Astrophys. J., {\bf636},
L117 (2006).
\bibitem{redshifterror}S. Bhattacharyya, M. C. Miller and F. K. Lamb, Astrophys. J.,
{\bf644}, 1085 (2006).
\bibitem{higherredshift}A. Tiengo, D. K. Galloway and T. di Salvo, et al., A\&A, {\bf441}, 283 (2005).
\bibitem{rotation}B. D. Lackey, M. Nayyar and B. J. Owen, Phys. Rev. D, {\bf73}, 024021
(2006).
\end{thebibliography}
\end{document}